\documentclass[preprint,superscriptaddress,showkeys,showpacs,nofootinbib,byrevtex]{revtex4} 
\usepackage{graphicx}
\usepackage{epstopdf}
\begin{document}      
\preprint{YITP-07-89}
\preprint{PNU-NTG-13/2007}
\preprint{PNU-NuRI-13/2007}
\title{Magnetic susceptibility of QCD vacuum at finite density 
\\from the nonlocal chiral quark model}
\author{Seung-il Nam}
\email[E-mail: ]{sinam@yukawa.kyoto-u.ac.jp}
\affiliation{Yukawa Institute for Theoretical Physics (YITP), Kyoto
University, Kyoto 606-8502, Japan} 
\author{Hyun-Chul Kim}
\affiliation{Department of Physics, Inha University, Incheon 402-751, Korea} 
\date{\today}
\begin{abstract}
We present in this talk a recent investigation on the magnetic
susceptibility ($\chi$) of the QCD vacuum at finite density,  
utilizing the nonlocal chiral quark model from the instanton vacuum.
We take into account the nonzero current-quark mass ($m_q$) explicitly
to consider the effect of explicit flavor SU(3) symmetry breaking.  It
turns out that, when we turn on the current-quark
mass ($m_q\ne0$), $\chi$ becomes smaller, indicating less response to the externally induced electromagnetic field, in comparison to that for $m_q=0$. 
\end{abstract} 
\maketitle
\section{Introduction}
Properties of hadrons are known to be changed at finite density
($\mu$) and/or temperature ($T$).  Thus, it is essential to understand
the mechanism of the modification of haronic properties in medium,
based on quantum chromodynamics (QCD) which is the underlying theory
of the strong interaction.  The first step to describe hadrons at
finite density is to investigate the in-medium modification of the QCD
vacuum.  

In the present talk, we want to study the magnetic susceptibility of
the QCD vacuum at the finite quark (number) chemical potential
($\mu_q$) at $T=0$, which is defined as 
\begin{equation}
\label{eq:VEV}
\langle{iq}^{\dagger}\sigma_{\mu\nu}q\rangle_{F}
=ie_q{F_{\mu\nu}}\langle{iq}^{\dagger}q\rangle\chi,
\end{equation}
where $\langle{i q}^{\dagger}q\rangle$ stands for the chiral
condensate. In principle, the magnetic susceptibility represents the
response of the vacuum in the presence of the external electromagnetic
(EM) field.  Moreover, the present study may shed light on the
interpretation for the magnetic properties of the compact star from 
the hadronic origin~\cite{Tatsumi:2005ri}.  

Here, we restrict ourselves to the spontaneous chiral symmetry
breaking (S$\chi$SB) phase.  We also take into account the nonzero
current-quark 
mass~\cite{Musakhanov:1998wp,Musakhanov:2001pc,Nam:2006ng}.  We
observe that, due to the competition between $m_q$ and $\mu_q$, the
finite current-quark mass changes the magnetic susceptibility
drastically, in particular, in the region of higher chemical
potentials, compared to the case of the chiral limit:
The $\chi_{\mu}$ for the massless quark increases stiffly  whereas it
is relatively flat or decreases with the finite current-quark mass.

\section{Nonlocal chiral quark model at finite density}
We start with the quark zero-mode equation with the
chemical potential in the (anti)instanton effects in Euclidean space: 
\begin{equation}
\label{eq:ZM}
\left[i\rlap{/}{\partial}-i\rlap{/}{\mu}+im_q+\rlap{/}{A}_{I\bar{I}}
\right]\Psi^{(0)}_{I\bar{I}}  =0.
\end{equation}
Note that the quark chemical potential four vector $\mu=(0,0,0,\mu_q)$
and current-quark mass $m_q$ are explicitly included in the
operator. $A_{\bar{I}I}$ stands for the (anti)instanton contribution
whereas the $\Psi^{(0)}_{I\bar{I}}$ denotes the quark zero-mode
solution. From this, the effective quark propagator in the
(anti)instanton effects can be derived as follows: 
\begin{equation}
\label{eq:aa1}
S=\frac{1}{i\rlap{/}{\partial}-i\rlap{/}{\mu}+im_q+i{\cal
    M}}, 
\,\,\,\,
{\cal M}\approx\hat{\cal M}(i\partial,\mu)
\left[\sqrt{1+\frac{m^2_q}{d}}-\frac{m_q}{d}\right],
\end{equation}
where ${\cal M}(i\partial,\mu)$ denotes the momentum-dependent
quark mass, which arises from the Fourier transformed zero-mode
solution of Eq.~(\ref{eq:ZM}), and its explicit form is given in
Ref.~\cite{Carter:1998ji}.  The parameter $d$ is chosen to be the
vacuum value, $198$ MeV for simplicity.  As for the effective quark
mass, we, however, parameterize it in the following form:
\begin{eqnarray}
\label{eq:MFD}
\hat{\cal M}(i\partial,\mu)&=&{\cal M}_0
\left[\frac{2\Lambda^2}{(i\partial-i\mu)(i\partial-i\mu)+2\Lambda^2}
\right]^2={\cal M}_0{\cal F}^2(i\partial,\mu),
\end{eqnarray}
for simplicity.  Note that this dynamical quark mass is complex in
general for $\mu_q\ne0$.  Here we use $\Lambda\approx600$ MeV and
${\cal M}_0\approx350$ MeV in connection to the instanton packing
fraction at vacuum, $(N/V)^{-4}\approx200$ MeV. Note that this parameterization works well in relatively dilute region, $\mu_\mathrm{ch.}\lesssim200$ MeV.

Now we are in a position to consider the vacuum expectation value
(VEV) shown in Eq.~(\ref{eq:VEV}) with the modified effective action,
derived from the instanton vacuum in the presence of the $\mu$ and EM
field: 
\begin{eqnarray}
\label{eq:GF}
{\cal S}_{\rm eff}[\mu,m_q,\mathcal{T}]=-{\rm Sp}\ln\left[
i\rlap{/}{D}-i\rlap{/}{\mu}+im_q+i{\cal M}(iD,\mu,m_q)+\sigma\cdot
\mathcal{T}\right], 
\end{eqnarray}
where ${\rm Sp}$ is the functional trace $\int d^4x\langle
x|\cdots|x\rangle$ running over Dirac and color indices. The $iD$
indicates the covariant derivative, $i\partial+e_qA$, in which $e_q$
and $A$ are the quark electric charge and externally induced photon
field, respectively.  $\mathcal{T}$ denotes the external tensor field.   
Performing the functional differentiation with respect to
$\mathcal{T}$, we obtain the following expression for the VEV of  
Eq.~(\ref{eq:VEV}) in a operator form: 
\begin{eqnarray}
\label{eq:VEV1}
-N_c{\rm tr}_{\gamma}\left\{\left[\frac{i\rlap{/}{D}-i\rlap{/}{\mu}
-i\bar{{\cal M}}_F}
{(i\partial-i\mu)^2+\frac{\sigma\cdot F}{2}-[i\rlap{/}{D},
{\cal M}_F]+\bar{{\cal M}}_F^2}
-\frac{i\rlap{/}{D}-i\rlap{/}{\mu}-im_q}
{-\partial^2+\frac{\sigma\cdot F}{2}+m_q^2}\right]
\sigma_{\mu\nu}
\right\},
\end{eqnarray}
where we use the abbreviation $\bar{\cal M}_F={\cal
  M}(iD,\mu_q,m_q)+m_q$ for convenience. $F_{\mu\nu}$ is the field
strength tensor for the external photon field.  $\mathrm{tr}_\gamma$
denotes the trace over spin space.  Finally we arrive at the
expression for the $\chi_{\mu}{\langle{i q}^{\dagger}q\rangle}_{\mu}$
as follows:   
\begin{eqnarray}
\label{eq:VEV2}
4N_c\int^{\infty}_{\infty}\frac{d^4p}{(2\pi)^4}
\left[\frac{\bar{{\cal M}}}
{[(p+i\mu)^2+\bar{{\cal M}}^2]^2}
-\frac{m_q}{[p^2+m_q^2]^2}
-\frac{{\cal M}'\cdot(p+i\mu)}{[(p+i\mu)^2+\bar{{\cal M}}^2]^2}\right],
\end{eqnarray}
where ${\cal M}={\cal M}(p,\mu,m_q)$. Here ${\cal M}'$ represents the
derivative of ${\cal M}$ with respect to $p$.  
\section{Numerical results}
\begin{figure}[t]
\begin{tabular}{cc}
\includegraphics[width=7cm]{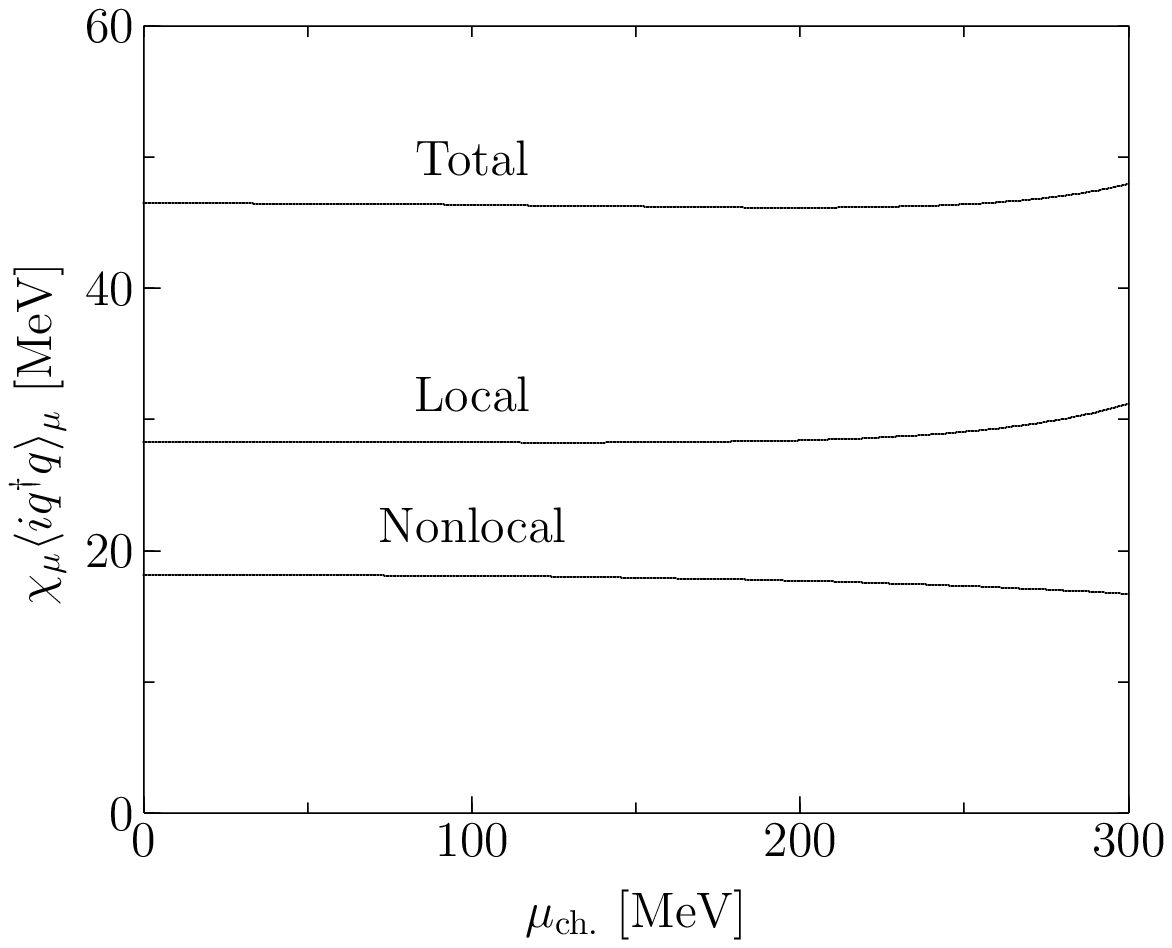}
\includegraphics[width=7cm]{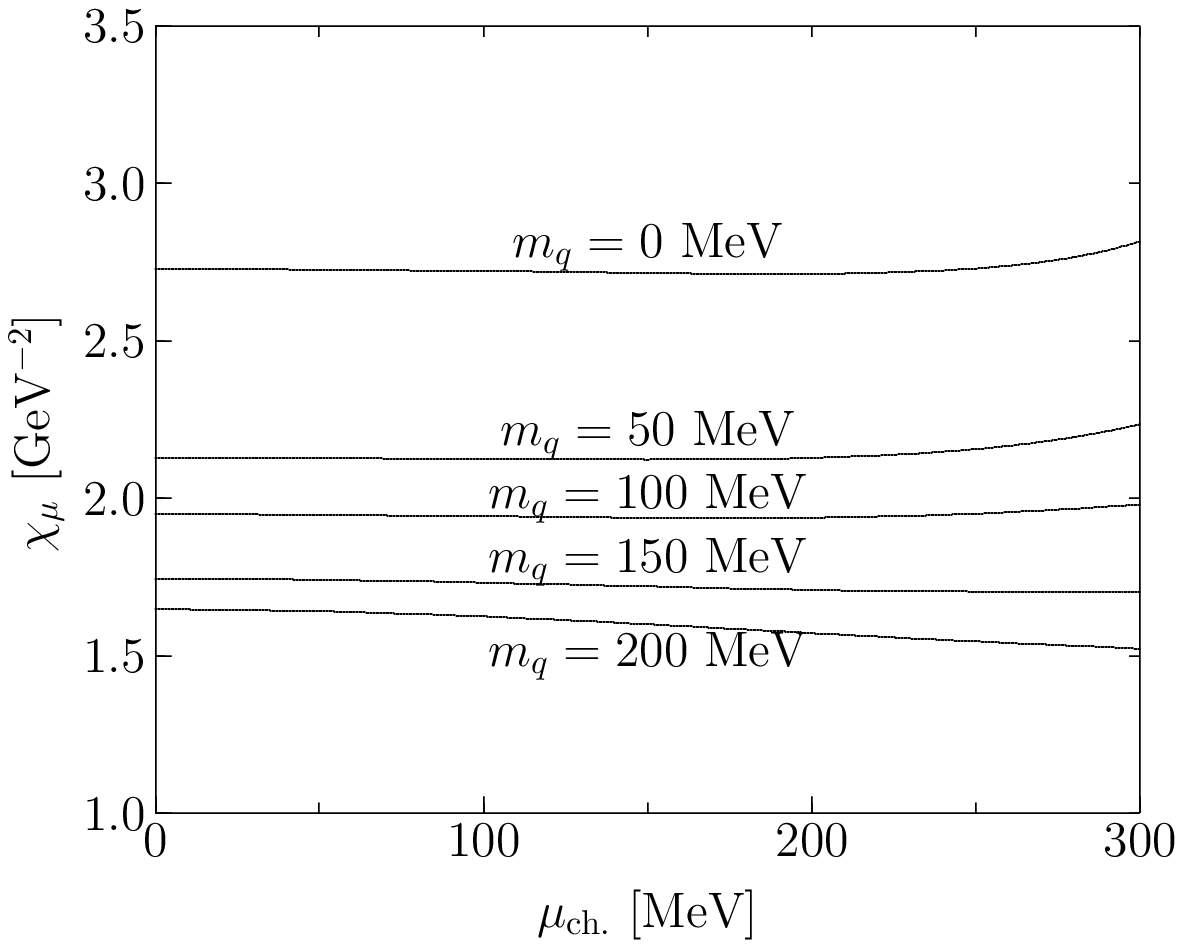}
\end{tabular}
\caption{Left panel: $\chi_{\mu}\langle iq^{\dagger}q\rangle_{\mu}$
  for each contribution. Right panel: $\chi_{\mu}$ for different values of the
  current-quark mass, $m_q$.}       
\label{fig1}
\end{figure}

First, those for the $\chi_{\mu}\langle iq^{\dagger}q\rangle_{\mu}$
are given in the left panel of Fig.~\ref{fig1} in which we draw it in
the chiral limit for each contribution separately. We draw them up to $\mu_\mathrm{ch.}\approx300$ MeV. Note that its
vacuum value becomes about $46.6$ MeV, which is consistent with that
computed in Ref.~\cite{Kim:2004hd}. This value is also compatible to
that from the 
QCD sum rules~\cite{Belyaev:1984ic,Balitsky:1985aq,Ball:2002ps}. While
the nonlocal contributions decrease slightly with
respect to the chemical potential, the local one increases slowly. This difference is due to the different 
behaviors of ${\cal M}$ and ${\cal M}'$ for the $\mu_q$.

In the right panel of Fig.~\ref{fig1}, we depict the $\chi_{\mu}$ for
different values of the $m_q$.  We note that the QCD magnetic susceptibilities computed are very stable with respect to $\mu_\mathrm{ch.}$. When we turn on the current-quark
mass ($m_q\ne0$), $\chi_{\mu}$ becomes smaller, indicating less response to the externally induced EM field.
\section{Summary and conclusion}
In the present talk, we have presented the magnetic susceptibility
($\chi_{\mu}$) of the QCD vacuum for the nonzero quark number chemical  
potential, $\mu_q\ne0$, based on the nonlocal chiral quark model for $N_f=1$,
derived from the instanton vacuum in the presence of the chemical
potential and external EM field.  We also have considered the effect
of flavor SU(3) symmetry breaking. We found that the QCD magnetic susceptibilities are in general very stable with respect to the chemical potential up to about $300$ MeV. When we turn on the current-quark mass, the magnetic susceptibility becomes smaller, in comparison to that for $m_q=0$.   

We conclude from the present results that the magnetic susceptibility
of the QCD vacuum at finite density depends much on the current-quark  
mass. These interesting observations may shed light on the extreme
phenomena taking place in the heavy-ion collision and compact stars,
in particular, associated with heavier quarks such as the strange
quark.  A detailed investigation is under progress and 
will appear elsewhere.  
\section*{Acknowledgments}
The authors are grateful to the organizers for the International
Workshop Chiral07, which was held during $13\sim16$ November, 2007 in
Osaka, Japan. The work of S.i.N. is partially supported by the grant
for Scientific Research (Priority Area No. 17070002) from the Ministry
of Education, Culture, Science and Technology, Japan. The work of
H.Ch.K. is supported by the Korea Research Foundation Grant funded by 
the Korean Government(MOEHRD) (KRF-2006-312-C00507).    

\end{document}